
%
%
\message{Welcome to the wonderful world of TeX}
\catcode`\@=11 
%
%
%
\font\seventeenrm=cmr17

\font\twelverm=cmr12
\font\ninerm=cmr9
\font\sixrm=cmr6

\font\seventeenbf=cmbx12 at 17pt
\font\fourteenbf=cmbx12 at 14pt
\font\twelvebf=cmbx12
\font\ninebf=cmbx9
\font\sixbf=cmbx6

\font\seventeeni=cmmi12 at 17pt             \skewchar\seventeeni='177
\font\fourteeni=cmmi12 at 14pt              \skewchar\fourteeni='177
\font\twelvei=cmmi12                        \skewchar\twelvei='177
\font\ninei=cmmi9                           \skewchar\ninei='177
\font\sixi=cmmi6                            \skewchar\sixi='177

\font\seventeensy=cmsy10 scaled\magstep3    \skewchar\seventeensy='60
\font\fourteensy=cmsy10 scaled\magstep2     \skewchar\fourteensy='60
\font\twelvesy=cmsy10 at 12pt               \skewchar\twelvesy='60
\font\ninesy=cmsy9                          \skewchar\ninesy='60
\font\sixsy=cmsy6                           \skewchar\sixsy='60

\font\seventeenex=cmex10 scaled\magstep3
\font\fourteenex=cmex10 scaled\magstep2
\font\twelveex=cmex10 at 12pt

\font\ninex=cmex9
\font\sevenex=cmex7
\font\sixex=cmex7 at 6pt
\font\fivex=cmex7 at 5pt

\font\seventeensl=cmsl10 scaled\magstep3
\font\fourteensl=cmsl10 scaled\magstep2
\font\twelvesl=cmsl10 scaled\magstep1
\font\ninesl=cmsl10 at 9pt
\font\sevensl=cmsl10 at 7pt
\font\sixsl=cmsl10 at 6pt
\font\fivesl=cmsl10 at 5pt

\font\seventeenit=cmti12 scaled\magstep2
\font\fourteenit=cmti12 scaled\magstep1
\font\twelveit=cmti12

\font\seventeentt=cmtt12 scaled\magstep2
\font\fourteentt=cmtt12 scaled\magstep1
\font\twelvett=cmtt12

\font\seventeencp=cmcsc10 scaled\magstep3
\font\fourteencp=cmcsc10 scaled\magstep2
\font\twelvecp=cmcsc10 scaled\magstep1
\font\tencp=cmcsc10

\newfam\cpfam

\font\seventeenbb=msbm10 scaled\magstep3
\font\fourteenbb=msbm10 scaled\magstep2
\font\twelvebb=msbm10 scaled\magstep1
\font\tenbb=msbm10
\font\ninebb=msbm9
\font\sevenbb=msbm7
\font\sixbb=msbm6
\font\fivebb=msbm5
\newfam\bbfam

\font\seventeenss=cmss17
\font\fourteenss=cmss12 at 14pt
\font\twelvess=cmss12
\font\tenss=cmss10
\font\niness=cmss9

\font\sevenss=cmss8 at 7pt
\font\sixss=cmss8 at 6pt
\font\fivess=cmss8 at 5pt
\newfam\ssfam

\font\seventeensxm=msam10 scaled\magstep3
\font\fourteensxm=msam10 scaled\magstep2
\font\twelvesxm=msam10 scaled\magstep1
\font\tensxm=msam10
\font\ninesxm=msam9
\font\sevensxm=msam7
\font\sixsxm=msam6
\font\fivesxm=msam5
\newfam\sxmfam

\font\seventeeneu=eufm10 at 17pt
\font\fourteeneu=eufm10 at 14pt
\font\twelveeu=eufm10 at 12pt
\font\teneu=eufm10
\font\nineeu=eufm9

\font\seveneu=eufm7
\font\sixeu=eufm6
\font\fiveeu=eufm5
\newfam\eufam

%
\def\hexnumber@#1{\ifnum#1<10 \number#1\else
 \ifnum#1=10 A\else\ifnum#1=11 B\else\ifnum#1=12 C\else
 \ifnum#1=13 D\else\ifnum#1=14 E\else\ifnum#1=15 F\fi\fi\fi\fi\fi\fi\fi}
\def\hexbb{\hexnumber@\bbfam}
\def\hexsxm{\hexnumber@\sxmfam}
\newdimen\b@gheight             \b@gheight=12pt
\newcount\f@ntkey               \f@ntkey=0
\def\f@m{\afterassignment\samef@nt\f@ntkey=}
\def\samef@nt{\fam=\f@ntkey \the\textfont\f@ntkey\relax}
\def\rm{\f@m0 }
\def\mit{\f@m1 }         
\def\cal{\f@m2 }
\def\it{\f@m\itfam}
\def\sl{\f@m\slfam}
\def\bf{\f@m\bffam}
\def\tt{\f@m\ttfam}
\def\caps{\f@m\cpfam}
\def\bb{\f@m\bbfam}
\def\ssf{\f@m\ssfam}
\def\sym{\f@m\sxmfam}
\def\eu{\f@m\eufam}
\def\seventeenpoint{\relax
    \textfont0=\seventeenrm          \scriptfont0=\twelverm
      \scriptscriptfont0=\ninerm
    \textfont1=\seventeeni           \scriptfont1=\twelvei
      \scriptscriptfont1=\ninei
    \textfont2=\seventeensy          \scriptfont2=\twelvesy
      \scriptscriptfont2=\ninesy
    \textfont3=\seventeenex          \scriptfont3=\twelveex
      \scriptscriptfont3=\ninex
    \textfont\itfam=\seventeenit    
    \textfont\slfam=\seventeensl    
      \scriptscriptfont\slfam=\ninesl
    \textfont\bffam=\seventeenbf     \scriptfont\bffam=\twelvebf
      \scriptscriptfont\bffam=\ninebf
    \textfont\ttfam=\seventeentt
    \textfont\cpfam=\seventeencp
    \textfont\bbfam=\seventeenbb     \scriptfont\bbfam=\twelvebb
      \scriptscriptfont\bbfam=\ninebb
    \textfont\ssfam=\seventeenss     \scriptfont\ssfam=\twelvess
      \scriptscriptfont\ssfam=\niness
    \textfont\sxmfam=\seventeensxm     \scriptfont\sxmfam=\twelvesxm
      \scriptscriptfont\sxmfam=\ninesxm
    \textfont\eufam=\seventeeneu     \scriptfont\eufam=\twelveeu
      \scriptscriptfont\eufam=\nineeu
    \samef@nt
    \b@gheight=17pt
    \setbox\strutbox=\hbox{\vrule height 0.85\b@gheight
                                depth 0.35\b@gheight width\z@ }}
\def\fourteenpoint{\relax
    \textfont0=\fourteencp          \scriptfont0=\tenrm
      \scriptscriptfont0=\sevenrm
    \textfont1=\fourteeni           \scriptfont1=\teni
      \scriptscriptfont1=\seveni
    \textfont2=\fourteensy          \scriptfont2=\tensy
      \scriptscriptfont2=\sevensy
    \textfont3=\fourteenex          \scriptfont3=\twelveex
      \scriptscriptfont3=\tenex
    \textfont\itfam=\fourteenit     \scriptfont\itfam=\tenit
    \textfont\slfam=\fourteensl     \scriptfont\slfam=\tensl
      \scriptscriptfont\slfam=\sevensl
    \textfont\bffam=\fourteenbf     \scriptfont\bffam=\tenbf
      \scriptscriptfont\bffam=\sevenbf
    \textfont\ttfam=\fourteentt
    \textfont\cpfam=\fourteencp
    \textfont\bbfam=\fourteenbb     \scriptfont\bbfam=\tenbb
      \scriptscriptfont\bbfam=\sevenbb
    \textfont\ssfam=\fourteenss     \scriptfont\ssfam=\tenss
      \scriptscriptfont\ssfam=\sevenss
    \textfont\sxmfam=\fourteensxm     \scriptfont\sxmfam=\tensxm
      \scriptscriptfont\sxmfam=\sevensxm
    \textfont\eufam=\fourteeneu     \scriptfont\eufam=\teneu
      \scriptscriptfont\eufam=\seveneu
    \samef@nt
    \b@gheight=14pt
    \setbox\strutbox=\hbox{\vrule height 0.85\b@gheight
                                depth 0.35\b@gheight width\z@ }}
\def\twelvepoint{\relax
    \textfont0=\twelverm          \scriptfont0=\ninerm
      \scriptscriptfont0=\sixrm
    \textfont1=\twelvei           \scriptfont1=\ninei
      \scriptscriptfont1=\sixi
    \textfont2=\twelvesy           \scriptfont2=\ninesy
      \scriptscriptfont2=\sixsy
    \textfont3=\twelveex          \scriptfont3=\ninex
      \scriptscriptfont3=\sixex
    \textfont\itfam=\twelveit    
    \textfont\slfam=\twelvesl    
      \scriptscriptfont\slfam=\sixsl
    \textfont\bffam=\twelvebf     \scriptfont\bffam=\ninebf
      \scriptscriptfont\bffam=\sixbf
    \textfont\ttfam=\twelvett
    \textfont\cpfam=\twelvecp
    \textfont\bbfam=\twelvebb     \scriptfont\bbfam=\ninebb
      \scriptscriptfont\bbfam=\sixbb
    \textfont\ssfam=\twelvess     \scriptfont\ssfam=\niness
      \scriptscriptfont\ssfam=\sixss
    \textfont\sxmfam=\twelvesxm     \scriptfont\sxmfam=\ninesxm
      \scriptscriptfont\sxmfam=\sixsxm
    \textfont\eufam=\twelveeu     \scriptfont\eufam=\nineeu
      \scriptscriptfont\eufam=\sixeu
    \samef@nt
    \b@gheight=12pt
    \setbox\strutbox=\hbox{\vrule height 0.85\b@gheight
                                depth 0.35\b@gheight width\z@ }}
\def\tenpoint{\relax
    \textfont0=\tenrm          \scriptfont0=\sevenrm
      \scriptscriptfont0=\fiverm
    \textfont1=\teni           \scriptfont1=\seveni
      \scriptscriptfont1=\fivei
    \textfont2=\tensy          \scriptfont2=\sevensy
      \scriptscriptfont2=\fivesy
    \textfont3=\tenex          \scriptfont3=\sevenex
      \scriptscriptfont3=\fivex
    \textfont\itfam=\tenit     \scriptfont\itfam=\seveni
    \textfont\slfam=\tensl     \scriptfont\slfam=\sevensl
      \scriptscriptfont\slfam=\fivesl
    \textfont\bffam=\tenbf     \scriptfont\bffam=\sevenbf
      \scriptscriptfont\bffam=\fivebf
    \textfont\ttfam=\tentt
    \textfont\cpfam=\tencp
    \textfont\bbfam=\tenbb     \scriptfont\bbfam=\sevenbb
      \scriptscriptfont\bbfam=\fivebb
    \textfont\ssfam=\tenss     \scriptfont\ssfam=\sevenss
      \scriptscriptfont\ssfam=\fivess
    \textfont\sxmfam=\tensxm     \scriptfont\sxmfam=\sevensxm
      \scriptscriptfont\sxmfam=\fivesxm
    \textfont\eufam=\teneu     \scriptfont\eufam=\seveneu
      \scriptscriptfont\eufam=\fiveeu
    \samef@nt
    \b@gheight=10pt
    \setbox\strutbox=\hbox{\vrule height 0.85\b@gheight
                                depth 0.35\b@gheight width\z@ }}
%
%
%
\normalbaselineskip = 15pt plus 0.2pt minus 0.1pt 
\normallineskip = 1.5pt plus 0.1pt minus 0.1pt
\normallineskiplimit = 1.5pt
\newskip\normaldisplayskip
\normaldisplayskip = 15pt plus 5pt minus 10pt 
\newskip\normaldispshortskip
\normaldispshortskip = 6pt plus 5pt
\newskip\normalparskip
\normalparskip = 6pt plus 2pt minus 1pt
\newskip\skipregister
\skipregister = 5pt plus 2pt minus 1.5pt
\newif\ifsingl@    \newif\ifdoubl@
\newif\iftwelv@    \twelv@true
\def\singlespace{\singl@true\doubl@false\spaces@t}
\def\doublespace{\singl@false\doubl@true\spaces@t}
\def\normalspace{\singl@false\doubl@false\spaces@t}
\def\Tenpoint{\tenpoint\twelv@false\spaces@t}
\def\Twelvepoint{\twelvepoint\twelv@true\spaces@t}
\def\spaces@t{\relax
      \iftwelv@ \ifsingl@\subspaces@t3:4;\else\subspaces@t1:1;\fi
       \else \ifsingl@\subspaces@t3:5;\else\subspaces@t4:5;\fi \fi
      \ifdoubl@ \multiply\baselineskip by 5
         \divide\baselineskip by 4 \fi }
\def\subspaces@t#1:#2;{
      \baselineskip = \normalbaselineskip
      \multiply\baselineskip by #1 \divide\baselineskip by #2
      \lineskip = \normallineskip
      \multiply\lineskip by #1 \divide\lineskip by #2
      \lineskiplimit = \normallineskiplimit
      \multiply\lineskiplimit by #1 \divide\lineskiplimit by #2
      \parskip = \normalparskip
      \multiply\parskip by #1 \divide\parskip by #2
      \abovedisplayskip = \normaldisplayskip
      \multiply\abovedisplayskip by #1 \divide\abovedisplayskip by #2
      \belowdisplayskip = \abovedisplayskip
      \abovedisplayshortskip = \normaldispshortskip
      \multiply\abovedisplayshortskip by #1
        \divide\abovedisplayshortskip by #2
      \belowdisplayshortskip = \abovedisplayshortskip
      \advance\belowdisplayshortskip by \belowdisplayskip
      \divide\belowdisplayshortskip by 2
      \smallskipamount = \skipregister
      \multiply\smallskipamount by #1 \divide\smallskipamount by #2
      \medskipamount = \smallskipamount \multiply\medskipamount by 2
      \bigskipamount = \smallskipamount \multiply\bigskipamount by 4 }
\def\normalbaselines{ \baselineskip=\normalbaselineskip
   \lineskip=\normallineskip \lineskiplimit=\normallineskip
   \iftwelv@\else \multiply\baselineskip by 4 \divide\baselineskip by 5
     \multiply\lineskiplimit by 4 \divide\lineskiplimit by 5
     \multiply\lineskip by 4 \divide\lineskip by 5 \fi }
\Twelvepoint  
%
\interlinepenalty=50
\interfootnotelinepenalty=5000
\predisplaypenalty=9000
\postdisplaypenalty=500
\hfuzz=1pt
\vfuzz=0.2pt
\dimen\footins=24 truecm 
\hoffset=10.5truemm 
\voffset=-8.5 truemm 
%
%
%
%
%
%
\def\footnote#1{\edef\@sf{\spacefactor\the\spacefactor}#1\@sf
      \insert\footins\bgroup\singl@true\doubl@false\Tenpoint
      \interlinepenalty=\interfootnotelinepenalty \let\par=\endgraf
        \leftskip=\z@skip \rightskip=\z@skip
        \splittopskip=10pt plus 1pt minus 1pt \floatingpenalty=20000
        \smallskip\item{#1}\bgroup\strut\aftergroup\@foot\let\next}
\skip\footins=\bigskipamount 
\dimen\footins=24truecm 
\newcount\fnotenumber
\def\clearfnotenumber{\fnotenumber=0}
\def\fnote{\advance\fnotenumber by1 \footnote{$^{\the\fnotenumber}$}}
\clearfnotenumber
%
%
\newcount\secnumber
\newcount\appnumber
\newif\ifs@c 
\newif\ifs@cd 
\s@cdtrue 
\def\unsectioned{\s@cdfalse\let\section=\subsection}
\def\clearappnumber{\appnumber=64}
\def\clearsecnumber{\secnumber=0}
\newskip\sectionskip         \sectionskip=\medskipamount
\newskip\headskip            \headskip=8pt plus 3pt minus 3pt
\newdimen\sectionminspace    \sectionminspace=10pc
\newdimen\referenceminspace  \referenceminspace=25pc
\def\Titlestyle#1{\par\begingroup \interlinepenalty=9999
     \leftskip=0.02\hsize plus 0.23\hsize minus 0.02\hsize
     \rightskip=\leftskip \parfillskip=0pt
     \advance\baselineskip by 0.5\baselineskip
     \hyphenpenalty=9000 \exhyphenpenalty=9000
     \tolerance=9999 \pretolerance=9000
     \spaceskip=0.333em \xspaceskip=0.5em
     \iftwelv@\seventeenpoint\else\fourteenpoint\fi
   \noindent #1\par\endgroup }
\def\titlestyle#1{\par\begingroup \interlinepenalty=9999
     \leftskip=0.02\hsize plus 0.23\hsize minus 0.02\hsize
     \rightskip=\leftskip \parfillskip=0pt
     \hyphenpenalty=9000 \exhyphenpenalty=9000
     \tolerance=9999 \pretolerance=9000
     \spaceskip=0.333em \xspaceskip=0.5em
     \iftwelv@\fourteenpoint\else\twelvepoint\fi
   \noindent #1\par\endgroup }
%
\def\spacecheck#1{\dimen@=\pagegoal\advance\dimen@ by -\pagetotal
   \ifdim\dimen@<#1 \ifdim\dimen@>0pt \vfil\break \fi\fi}
\def\section#1{\cleareqnumber \s@ctrue \global\advance\secnumber by1
   \message{Section \the\secnumber: #1}
   \par \ifnum\the\lastpenalty=30000\else
   \penalty-200\vskip\sectionskip \spacecheck\sectionminspace\fi
   \noindent {\caps\enspace\S\the\secnumber\quad #1}\par
   \nobreak\vskip\headskip \penalty 30000 }
\def\subsection#1{\par
   \ifnum\the\lastpenalty=30000\else \penalty-100\smallskip
   \spacecheck\sectionminspace\fi
   \noindent\undertext{#1}\enspace \vadjust{\penalty5000}}

\def\undertext#1{\vtop{\hbox{#1}\kern 1pt \hrule}}
\def\subsubsection#1{\par
   \ifnum\the\lastpenalty=30000\else \penalty-100\smallskip \fi
   \noindent\hbox{#1}\enspace \vadjust{\penalty5000}}

\def\appendix#1{\cleareqnumber \s@cfalse \global\advance\appnumber by1
   \message{Appendix \char\the\appnumber: #1}
   \par \ifnum\the\lastpenalty=30000\else
   \penalty-200\vskip\sectionskip \spacecheck\sectionminspace\fi
   \noindent {\caps\enspace Appendix \char\the\appnumber\quad #1}\par
   \nobreak\vskip\headskip \penalty 30000 }
\clearsecnumber
\clearappnumber
%
%
\def\ack{\par\penalty-100\medskip \spacecheck\sectionminspace
   \line{\iftwelv@\fourteencp\else\twelvecp\fi\hfil ACKNOWLEDGEMENTS\hfil}%
\nobreak\vskip\headskip }
\def\refs{\begingroup \par\penalty-100\medskip \spacecheck\sectionminspace
   \line{\iftwelv@\fourteencp\else\twelvecp\fi\hfil REFERENCES\hfil}%
\nobreak\vskip\headskip \frenchspacing }
\def\endrefs{\par\endgroup}
%
\newcount\refnumber
\def\clearrefnumber{\refnumber=0}  \clearrefnumber
\newwrite\R@fs                              
\immediate\openout\R@fs=\jobname.references 
\def\closerefs{\immediate\closeout\R@fs} 
\def\refsout{\closerefs\refs
\catcode`\@=11                          
\input\jobname.references               
\catcode`\@=12			        
\endrefs}
\def\refitem#1{\item{{\bf #1}}}
\def\ifundefined#1{\expandafter\ifx\csname#1\endcsname\relax}
%
%
\def\[#1]{\ifundefined{#1R@FNO}%
\global\advance\refnumber by1%
\expandafter\xdef\csname#1R@FNO\endcsname{[\the\refnumber]}%
\immediate\write\R@fs{\noexpand\refitem{\csname#1R@FNO\endcsname}%
\noexpand\csname#1R@F\endcsname}\fi{\bf \csname#1R@FNO\endcsname}}
\def\refdef[#1]#2{\expandafter\gdef\csname#1R@F\endcsname{{#2}}}
%
%
%
%
%
%
\newcount\eqnumber
\def\cleareqnumber{\eqnumber=0}
\newif\ifal@gn \al@gnfalse  
\def\veqnalign#1{\al@gntrue \vbox{\eqalignno{#1}} \al@gnfalse}
\def\eqnalign#1{\al@gntrue \eqalignno{#1} \al@gnfalse}
\def\(#1){\relax%
\ifundefined{#1@Q}
 \global\advance\eqnumber by1
 \ifs@cd
  \ifs@c
   \expandafter\xdef\csname#1@Q\endcsname{{%
\noexpand\rm(\the\secnumber .\the\eqnumber)}}
  \else
   \expandafter\xdef\csname#1@Q\endcsname{{%
\noexpand\rm(\char\the\appnumber .\the\eqnumber)}}
  \fi
 \else
  \expandafter\xdef\csname#1@Q\endcsname{{\noexpand\rm(\the\eqnumber)}}
 \fi
 \ifal@gn
    & \csname#1@Q\endcsname
 \else
    \eqno \csname#1@Q\endcsname
 \fi
\else%
\csname#1@Q\endcsname\fi\global\let\@Q=\relax}
%
%
\newif\iffrontpage \frontpagefalse
\newif\ifletterstyle
\newif\ifmemostyle
\newif\ifpoemstyle
\newif\ifspanish \spanishfalse
\newtoks\memofootline
\newtoks\memoheadline
\newtoks\paperfootline
\newtoks\letterfootline
\newtoks\paperheadline
\newtoks\letterheadline
\newtoks\date
\newtoks\fecha
\newtoks\d@te
\headline={\ifmemostyle\the\memoheadline%
           \else \ifletterstyle\the\letterheadline%
           \else \ifpoemstyle\hfil%
           \else \the\paperheadline\fi\fi\fi}
\paperheadline={\hfil}
\letterheadline={\ifnum\pageno=1 \hfil
  \else\ifspanish\rm p\'agina \ \folio\hfil\the\fecha
  \else\rm page \ \folio\hfil\the\date\fi\fi}
\memoheadline={\ss Page \ \folio\hfil\the\date}
\footline={\ifmemostyle\the\memofootline\else\ifpoemstyle\hfil\else%
           \ifletterstyle\the\letterfootline\else\the\paperfootline\fi\fi\fi}
\letterfootline={\hfil}
\memofootline={\hfil}
\paperfootline={\iffrontpage\hfil\else \hss\iftwelv@\twelverm\else\tenrm\fi
-- \folio\ --\hss \fi }
\def\monthname{\relax\ifcase\month 0/\or January\or February\or
   March\or April\or May\or June\or July\or August\or September\or
   October\or November\or December\else\number\month/\fi}
\def\today{\monthname\ \number\day, \number\year}
\date={\today}
\def\nombremes{\relax\ifcase\month 0/\or Enero\or Febrero\or
   Marzo\or Abril\or Mayo\or Junio\or Julio\or Agosto\or Septiembre\or
   Octubre\or Noviembre\or Diciembre\else\number\month/\fi}
\def\hoy{\number\day\ de\ \nombremes, \number\year}
\fecha={\hoy}
\def\itpletterhead{
       \def\k{\kern -3pt}
       \font\sblogo=cmbx12 scaled \magstep5
       \setbox1=\hbox{\sblogo S\k t\k o\k n\k y\k B\k r\k o\k o\k k\k}
       \setbox2=\hbox{\vbox{
                      \hbox{\tenrm Jos\'e M. Figueroa-O'Farrill}
                      \hbox{\tenrm Institute for Theoretical Physics}
                      \hbox{\tenrm State University of New York at Stony Brook}
                      \hbox{\tenrm Stony Brook, NY\ \ 11794--3840}
                      \hbox{\tenrm telephone: (516) 632-7965}}}
\hbox to \hsize{\vbox to 1.2in{\vfill\box1\vfill}\hfill
                \vbox to 1.2in{\box2\vfill}}}
\hyphenation{U-ni-ver-si-teit u-ni-ver-si-teit The-o-re-tische the-o-re-tische
Fy-si-ca fy-si-ca}
\def\kulletterhead{
       \setbox1=\hbox{\vbox{
                      \hbox{\tenrm Jos\'e M. Figueroa-O'Farrill}
                      \hbox{\tenrm Instituut voor Theoretische Fysica}
                      \hbox{\tenrm Katholieke Universiteit Leuven}
                      \hbox{\tenrm Celestijnenlaan 200 D}
                      \hbox{\tenrm B-3001 Heverlee, BELGIUM}
                      \hbox{\tentt e-mail: fgbda11@blekul11.bitnet}
                      \hbox{\tenrm telephone: (016) 201015 x3219}}}
\hbox to \hsize{\hfill\vbox to 1.5truein{\box1\vfill}}}
\def\bonnletterhead{
       \setbox1=\hbox{\vbox{
                      \hbox{\tenrm Jos\'e M. Figueroa-O'Farrill}
                      \hbox{\tenrm Physikalisches Institut der}
                      \hbox{\tenrm Universit\"at Bonn}
                      \hbox{\tenrm Nussallee 12}
                      \hbox{\tenrm W-5300 Bonn 1, GERMANY}
                      \hbox{\tentt e-mail: figueroa@pib1.physik.uni-bonn.de}
                      \hbox{\tenrm telephone: (0228)73-2224}}}
\hbox to \hsize{\hfill\vbox to 1.5truein{\box1\vfill}}}
\def\jsletterhead{
       \setbox1=\hbox{\vbox{
                      \hbox{\tenrm Jos\'e M. Figueroa-O'Farrill}
                      \hbox{\tenrm Stany Schrans}
                      \hbox{\tenrm Instituut voor Theoretische Fysica}
                      \hbox{\tenrm Katholieke Universiteit Leuven}
                      \hbox{\tenrm Celestijnenlaan 200 D}
                      \hbox{\tenrm B-3001 Heverlee, BELGIUM}
                      \hbox{\tenrm e-mail: fgbda31@blekul11.bitnet}}}
\hbox to \hsize{\hfill\vbox to 1.5truein{\box1\vfill}}}
\def\jeletterhead{
       \setbox1=\hbox{\vbox{
                      \hbox{\tenrm Jos\'e M. Figueroa-O'Farrill}
                      \hbox{\tenrm Eduardo Ramos}
                      \hbox{\tenrm Instituut voor Theoretische Fysica}
                      \hbox{\tenrm Katholieke Universiteit Leuven}
                      \hbox{\tenrm Celestijnenlaan 200 D}
                      \hbox{\tenrm B-3001 Heverlee, BELGIUM}
                      \hbox{\tenrm e-mail: fgbda11@blekul11.bitnet}}}
\hbox to \hsize{\hfill\vbox to 1.5truein{\box1\vfill}}}
\def\Date{\ifspanish\d@te=\fecha\else\d@te=\date\fi
\line{\hfill\rm\the\d@te}\bigskip}

%

%

%
\def\paperstyle{\letterstylefalse\normalspace\papersize}
\def\letterstyle{\letterstyletrue\singlespace\lettersize\parindent=0pt
                 \advance\parskip by 2\parskip}
\def\storystyle{\letterstyletrue\singlespace\lettersize
                 \advance\parskip by 1.5\parskip}

\def\papersize{\hsize=14 truecm\vsize=22 truecm
               \skip\footins=\bigskipamount}
\def\lettersize{\hsize=14truecm\vsize=22truecm
   \skip\footins=\smallskipamount \multiply\skip\footins by 3 }
\paperstyle   
%
%
%
\newskip\frontpageskip
\newif\ifp@bblock \p@bblocktrue
\newif\ifm@nth \m@nthtrue
\newtoks\pubnum
\newtoks\pubtype
\newtoks\m@nthn@me
\newcount\Ye@r
\advance\Ye@r by \year
\advance\Ye@r by -1900
\def\Year#1{\Ye@r=#1}
\def\Month#1{\m@nthfalse \m@nthn@me={#1}}
\def\m@nthname{\ifm@nth\monthname\else\the\m@nthn@me\fi}
\def\titlepage{\global\frontpagetrue\paperstyle\hrule height\z@ \relax
               \ifp@bblock\pubblock\fi\relax }
\def\endtitlepage{\vfil\break
                  \frontpagefalse} 
\frontpageskip=12pt plus .5fil minus 2pt
\pubtype={\iftwelv@\twelvesl\else\tensl\fi\ (Preliminary Version)}
\pubnum={?}
\def\nopubblock{\p@bblockfalse}
\def\pubblock{\line{\hfil\iftwelv@\twelverm\else\tenrm\fi%
BONN--HE--\number\Ye@r--\the\pubnum\the\pubtype}
              \line{\hfil\iftwelv@\twelverm\else\tenrm\fi%
\m@nthname\ \number\year}}
\def\title#1{\vskip\frontpageskip\Titlestyle{\caps #1}\vskip3\headskip}
\def\author#1{\vskip.5\frontpageskip\titlestyle{\caps #1}\nobreak}
\def\andauthor{\vskip.5\frontpageskip\centerline{and}\author}

\def\address#1{\par\kern 5pt\titlestyle{
\it #1}}
\def\andaddress{\par\kern 5pt \centerline{\sl and} \address}

\def\KUL{\address{Instituut voor Theoretische Fysica\break
                  Universiteit Leuven\break
                  Celestijnenlaan 200 D\break
                  B--3001 Heverlee, BELGIUM}}
\def\Bonn{\address{Physikalisches Institut der\break
                   Universit\"at Bonn\break
                   Nu{\ss}allee 12\break
                   W--5300 Bonn 1, GERMANY}}
\def\abstract#1{\par\dimen@=\prevdepth \hrule height\z@ \prevdepth=\dimen@
   \vskip\frontpageskip\spacecheck\sectionminspace
   \centerline{\iftwelv@\fourteencp\else\twelvecp\fi ABSTRACT}\vskip\headskip
   {\noindent #1}}
%

%
%
%
\def\leaderfill{\leaders\hbox to 1em{\hss.\hss}\hfill}
\def\boxit#1{\vcenter{\hrule\hbox{\vrule\kern8pt
      \vbox{\kern8pt#1\kern8pt}\kern8pt\vrule}\hrule}}

%
%
%
\def\ref#1{{\bf [#1]}}
\def\ie{{\it i.e.\/}}
\def\eg{{\it e.g.\/}}
\def\th{{\rm th}}
\def\nl{\hfil\break}
%
%
%
%
%
\newif\ifm@thstyle \m@thstylefalse
\def\mathstyle{\m@thstyletrue}
\def\proclaim#1#2\par{\smallbreak\begingroup
\advance\baselineskip by -0.25\baselineskip%
\advance\belowdisplayskip by -0.35\belowdisplayskip%
\advance\abovedisplayskip by -0.35\abovedisplayskip%
    \noindent{\caps#1.\enspace}{#2}\par\endgroup%
\smallbreak}
\def\m@kem@th<#1>#2#3{%
\ifm@thstyle \global\advance\eqnumber by1
 \ifs@cd
  \ifs@c
   \expandafter\xdef\csname#1\endcsname{{%
\noexpand #2\ \the\secnumber .\the\eqnumber}}
  \else
   \expandafter\xdef\csname#1\endcsname{{%
\noexpand #2\ \char\the\appnumber .\the\eqnumber}}
  \fi
 \else
  \expandafter\xdef\csname#1\endcsname{{\noexpand #2\ \the\eqnumber}}
 \fi
 \proclaim{\csname#1\endcsname}{#3}
\else
 \proclaim{#2}{#3}
\fi}
%
%
%
%
%
%
\def\Thm<#1>#2{\m@kem@th<#1M@TH>{Theorem}{\sl#2}}
\def\Prop<#1>#2{\m@kem@th<#1M@TH>{Proposition}{\sl#2}}
\def\Def<#1>#2{\m@kem@th<#1M@TH>{Definition}{\rm#2}}
\def\Lem<#1>#2{\m@kem@th<#1M@TH>{Lemma}{\sl#2}}
\def\Cor<#1>#2{\m@kem@th<#1M@TH>{Corollary}{\sl#2}}
\def\Conj<#1>#2{\m@kem@th<#1M@TH>{Conjecture}{\sl#2}}
\def\Rmk<#1>#2{\m@kem@th<#1M@TH>{Remark}{\rm#2}}
\def\Exm<#1>#2{\m@kem@th<#1M@TH>{Example}{\rm#2}}
\def\Qry<#1>#2{\m@kem@th<#1M@TH>{Query}{\it#2}}
%
%

\def\<#1>{\csname#1M@TH\endcsname}
%
%

\def\lapprox{\hbox{\lower3pt\hbox{$\buildrel<\over\sim$}}}
\def\gapprox{\hbox{\lower3pt\hbox{$\buildrel<\over\sim$}}}
\def\quotient#1#2{#1/\lower0pt\hbox{${#2}$}}
%
%
\def\to{\rightarrow}
%

%
%
%
%
%
\def\underrightarrow#1{\vtop{\ialign{##\crcr
      $\hfil\displaystyle{#1}\hfil$\crcr
      \noalign{\kern-\p@\nointerlineskip}
      \rightarrowfill\crcr}}} 
\def\underleftarrow#1{\vtop{\ialign{##\crcr
      $\hfil\displaystyle{#1}\hfil$\crcr
      \noalign{\kern-\p@\nointerlineskip}
      \leftarrowfill\crcr}}}  

%
%
\def\comm#1#2{\bigl[#1\, ,\,#2\bigr]}
%
%
\def\pder#1#2{{{\partial #1}\over{\partial #2}}}
%
%
%
%
%
%
\newdimen\unit
\newdimen\redunit
%
%
\def\p@int#1:#2 #3 {\rlap{\kern#2\unit
     \raise#3\unit\hbox{#1}}}
%
%
\def\th@r{\vrule height0\unit depth.1\unit width1\unit}
\def\bh@r{\vrule height.1\unit depth0\unit width1\unit}
\def\lv@r{\vrule height1\unit depth0\unit width.1\unit}
\def\rv@r{\vrule height1\unit depth0\unit width.1\unit}
%
%
\def\t@ble@u{\hbox{\p@int\bh@r:0 0
                   \p@int\lv@r:0 0
                   \p@int\rv@r:.9 0
                   \p@int\th@r:0 1
                   }
             }
%
%
\def\t@bleau#1#2{\rlap{\kern#1\redunit
     \raise#2\redunit\t@ble@u}}
%
%
\newcount\n
\newcount\m
\def\makecol#1#2#3{\n=0 \m=#3
  \loop\ifnum\n<#1{}\advance\m by -1 \t@bleau{#2}{\number\m}\advance\n by 1
\repeat}
%
%
\def\makerow#1#2#3{\n=0 \m=#3
 \loop\ifnum\n<#1{}\advance\m by 1 \t@bleau{\number\m}{#2}\advance\n by 1
\repeat}
%
%
\def\checkunits{\ifinner \unit=6pt \else \unit=8pt \fi
                \redunit=0.9\unit } 
\def\ytsym#1{\checkunits\kern-.5\unit
  \vcenter{\hbox{\makerow{#1}{0}{0}\kern#1\unit}}\kern.5em} 
\def\ytant#1{\checkunits\kern.5em
  \vcenter{\hbox{\makecol{#1}{0}{0}\kern1\unit}}\kern.5em} 
\def\ytwo#1#2{\checkunits
  \vcenter{\hbox{\makecol{#1}{0}{0}\makecol{#2}{1}{0}\kern2\unit}}
                  \ } 
\def\ythree#1#2#3{\checkunits
  \vcenter{\hbox{\makecol{#1}{0}{0}\makecol{#2}{1}{0}\makecol{#3}{2}{0}%
\kern3\unit}}
                  \ } 
%
%
%

\def\NPB#1#2#3{{\sl Nucl. Phys.} {\bf B#1} (#2) #3}

\def\CMP#1#2#3{{\sl Comm. Math. Phys.} {\bf #1} (#2) #3}

\def\PLA#1#2#3{{\sl Phys. Lett.} {\bf #1A} (#2) #3}
\def\PLB#1#2#3{{\sl Phys. Lett.} {\bf #1B} (#2) #3}
\def\JMP#1#2#3{{\sl J. Math. Phys.} {\bf #1} (#2) #3}

\def\AoP#1#2#3{{\sl Ann. of Phys.} {\bf #1} (#2) #3}

\def\Invm#1#2#3{{\sl Invent. math.} {\bf #1} (#2) #3}

\def\IJMPA#1#2#3{{\sl Int. J. Mod. Phys.} {\bf A#1} (#2) #3}

\catcode`\@=12 
%
%
%
%
\def\d{\partial}
\let\pb=\anticomm
\def\pdo{{\hbox{$\Psi$DO}}}

\def\fr#1/#2{\ifinner{{\scriptstyle#1}\over{\scriptstyle#2}}\else%
                      {{#1}\over{#2}}\fi}

\def\Jt{\widetilde{J}}

\def\Diff{{\rm Diff}}
\def\dlb#1#2{\lbrack\!\lbrack#1,#2\rbrack\!\rbrack}

\def\kpder#1#2#3{{{\partial^{#1} #2}\over{\partial^{#1} #3}}}
\refdef[KW]{B.~A.~Kupershmidt and G.~Wilson, \Invm{62}{1981}{403}.}
\refdef[DickeyKW]{L.~A.~Dickey, \CMP{87}{1982}{127}.}
\refdef[Adler]{M.~Adler, \Invm{50}{1981}{403}.}
\refdef[GD]{I.~M.~Gel'fand and L.~A.~Dickey, {\sl A family of Hamiltonian
structures connected with integrable nonlinear differential equations},
Preprint 136, IPM AN SSSR, Moscow (1978).}
\refdef[Magri]{F.~Magri, \JMP{19}{1978}{1156}.}
\refdef[DFIZ]{P.~Di Francesco, C.~Itzykson, and J.-B.~Zuber,
\CMP{140}{1991}{543}.}
\refdef[WKP]{L.~A.~Dickey, {\sl Annals of the New York Academy of
Science} {\bf 491} (1987) 131;\nl
J.~M.~Figueroa-O'Farrill, J.~Mas, and E.~Ramos, \PLB{266}{1991}{298}.}
\refdef[Krich]{I. Krichever, \CMP{143}{1992}{415}.}
\refdef[DVV]{R. Dijkgraaf, H. Verlinde, and E. Verlinde,
\NPB{352}{1991}{59}.}
\refdef[TakaTake]{K. Takasaki and T. Takebe, {\sl SDIFF(2) KP
Hierarchy}, Preprint RIMS-814, December 1991.}
\refdef[FuKaNa]{M. Fukuma, H. Kawai, and R. Nakayama,
\IJMPA{6}{1991}{1385}.}
\refdef[KoGi]{Y. Kodama, \PLA{129}{1988}{223},
\PLA{147}{1990}{477};\nl
Y. Kodama and J. Gibbons, \PLA{135}{1989}{167}.}
\refdef[Laxform]{J.~M.~Figueroa-O'Farrill and E.~Ramos,
\PLB{262}{1991}{265}.}
\refdef[KoSt]{B. Kostant and S. Sternberg, \AoP{176}{1987}{49}.}
\refdef[Next]{J. M. Figueroa-O'Farrill and E. Ramos, {\sl }
Preprint--KUL--TF--92/6, February 1992, in preparation.}
\refdef[Uniw]{J. M. Figueroa-O'Farrill and E. Ramos, {\sl Existence
and Uniqueness of the Universal $W$-Algebra}, to appear in
the {\sl J.~Math.~Phys.}}
\refdef[Unpub]{J. M. Figueroa-O'Farrill and E. Ramos, {\sl An infinite
number of hamiltonian structures for the KP hierarchy}, unpublished.}
\refdef[DickeyPC]{L.~A.~Dickey, private communication.}
\refdef[Bakas]{I. Bakas, \CMP{134}{1990}{487}.}
\refdef[Pope]{C. N. Pope, L. J. Romans, and X. Shen,
\PLB{242}{1990}{401}.}
\overfullrule=0pt
\unsectioned
\def\pubblock{ \line{\hfil\twelverm Preprint--KUL--TF--92/5}
               \line{\hfil\twelverm BONN--HE--92--03}
               \line{\hfil\twelvett hepth@xxx/9202040}
               \line{\hfil\twelverm February 1992}}
\titlepage
\title{THE CLASSICAL LIMIT OF $W$--ALGEBRAS}
\author{Jos\'e M. Figueroa-O'Farrill\footnote{${}^\natural$}{\tt
e-mail: figueroa@pib1.physik.uni-bonn.de}}
\Bonn
\andauthor{Eduardo Ramos\footnote{${}^\sharp$}{\tt
e-mail: fgbda06@blekul11.bitnet}}
\KUL
\abstract{We define and compute explicitly the classical limit of the
realizations of $W_n$ appearing as hamiltonian structures of
generalized KdV hierarchies.  The classical limit is obtained by
taking the commutative limit of the ring of pseudodifferential
operators.  These algebras---denoted $w_n$---have free field
realizations in which the generators are given by the elementary
symmetric polynomials in the free fields.  We compute the algebras
explicitly and we show that they are all reductions of a new algebra
$w_{\rm KP}$, which is proposed as the universal classical $W$-algebra
for the $w_n$ series.  As a deformation of this algebra we also
obtain $w_{1+\infty}$, the classical limit of $W_{1+\infty}$.}
\endtitlepage
\section{Introduction}

It is by now well known that classical realizations of $W$-algebras
appear naturally as Poisson brackets of integrable hierarchies of Lax
type.  For example, the Virasoro algebra is realized as the Magri
bracket for the KdV hierarchy \[Magri], and the
Zamolodchikov-Fateev-Lykyanov $W_n$ algebras as the second
Gel'fand--Dickey bracket for the generalized $n^\th$-order KdV
hierarchy \[GD].  Although the $W$-algebras appearing in this fashion
are often unfortunately (mis)named ``classical'' $W$-algebras, we
prefer to refer to these algebras as {\sl classical realizations} of
$W$-algebras and reserve the name {\sl classical $W$-algebras} to the
nonlinear extensions of $\Diff\,S^1$ which will be the topic of
this paper.

We will obtain classical $W$-algebras by taking the classical limit of
the second Gel'fand--Dickey bracket: these are the classical limits of
the $W_n$-algebras and we denote them by $w_n$.  We can think of this
classical limit as a suitable contraction of the Gel'fand--Dickey
algebra, but more fundamentally as the algebra which is induced after
taking the classical (\ie, commutative) limit of the relevant
algebraic structures in the Lax formalism, namely the ring of
pseudodifferential operators.

The classical $W$-algebras we will obtain are, in a way, the simplest
examples of $W$-algebras.  The algebra $w_n$ is a nonlinear extension
of $\Diff\,S^1$ by primary fields of weights $3,4,\ldots,n$.  These
algebras are moreover easy to write down explicitly and have the
advantage that the natural fields generating the algebra are already
primaries.  They therefore seem to be unique candidates to investigate
questions (\eg, $W$-geometry) whose answers seem to be obscured by the
more complicated quantum or classical realizations of these algebras.

The purpose of this paper is to announce results whose proofs will be
given in a forthcoming publication \[Next] containing also numerous
extensions of these results; particularly in the supersymmetric
arena.  The plan of the paper is the following.  After briefly
reviewing the emergence of $W$-algebras in the Lax formalism, we
define their classical limit by first taking the classical limit of
the ring of pseudodifferential operators, and then using a classical
version of the Kupershmidt--Wilson theorem \[KW]\[DickeyKW].  In this
context, this theorem states that $w_n$ is (a reduction of) the
Poisson algebra of symmetric polynomials in $n$ free fields.  We then
write the algebras explicitly and show that the basis fields in which
we write the algebra are already primaries.  We then show that all
these algebras are reductions of $w_{\rm KP}$: the classical limit of
$W_{\rm KP}$---the ``natural'' second hamiltonian structure of the
KP hierarchy \[WKP].  Perhaps this needs some clarification.  In
\[WKP] it was shown that the KP hierarchy is bihamiltonian: the second
hamiltonian structure begin given by the natural generalization of the
Adler map to the space of KP operators and the first structure being
given by deforming this map as in \[KW].  However the KP hierarchy has
an infinite number of inequivalent bihamiltonian structures indexed by
the natural numbers, of which the one in \[WKP] is to some extent the
natural one.  The existence of this infinite number of bihamiltonian
structures was proven in \[Unpub] but appears to have been first known
to Radul \[DickeyPC].  We end the paper with a brief note on the
universal properties of $w_{\rm KP}$, its deformation to
$w_{1+\infty}$, and some concluding remarks.

The classical limit of the Lax formalism implicit in our
considerations yields integrable hierarchies known as the
``dispersionless KdV hierarchies'' which are reductions of the
Khokhlov-Zabolotskaya or dispersionless KP hierarchy.  These
hierarchies have been studied in \[KoGi] and \[TakaTake].  Particular
solutions of the dispersionless Lax equations have been shown in
\[Krich] to correspond to the perturbed chiral ring in topological
minimal models \[DVV].  The analogs to the $\tau$-functions for these
models should therefore obey $w_n$ constraints, obtained as reductions
of $w_{\rm KP}$ constraints as was done in \[FuKaNa] for the matrix
model formulation of $2d$-quantum gravity.
\section{$W$-algebras from Lax Operators}

In this section we briefly describe how $W$ algebras appear as Poisson
structures in the space of Lax operators.  Brevity demands that we
omit a review of the formalism, which can be found in the literature
in various amounts of detail---see, for example, \[Laxform].

Let $L=\d^n + \sum_{i=0}^{n-1}u_i\d^i$ be a differential operator and
$X= \sum_{i=0}^{n-1}\d^{-i-1}x_i$ be a pseudodifferential operator.
The Adler map $J$, defined by \[Adler]
$$J(X) \equiv (LX)_+L - L(XL)_+ = L(XL)_- - (LX)_-L~,\(Adler)$$
sends $X$ linearly to a differential operator of order at most $n-1$.
Therefore we can write
$$J(X) = \sum_{i,j=0}^{n-1} \left( J_{ij}\cdot x_j\right) \d^i~,\()$$
which defines differential operators $J_{ij}$.  Gel'fand and Dickey
\[GD] proved that the $J_{ij}$ define a Poisson bracket by
$$\pb{u_i(x)}{u_j(y)} = - J_{ij}\cdot \delta(x-y)~,\(gdbra)$$
where the operator $J_{ij}$ is taken at the point $x$.  This bracket
is known as the second Gel'fand--Dickey bracket and is the second
hamiltonian structure for the $n^\th$-order generalized KdV
hierarchy---the hierarchy of isospectral flows of the Lax operator
$L$.

A more conceptual proof of the fact that the bracket \(gdbra) is
Poisson derives from the Kupershmidt--Wilson theorem \[KW]\[DickeyKW].
This theorem states the following. Let us factorize the Lax operator
as follows:
$$L = \d^n + \sum_{i=0}^{n-1} u_i\d^i=
(\d+\phi_1)(\d+\phi_2)\cdots(\d+\phi_n)~.\()$$
This way the $\{u_i\}$ are given as differential polynomials of the
$\{\phi_i\}$.  If we then define
$$\pb{\phi_i(x)}{\phi_j(y)}=\delta_{ij}\delta'(x-y)~,\()$$
and compute the induced bracket of the $\{u_i(\phi)\}$, we find precisely
\(gdbra).

We can define another Poisson bracket as a deformation of \(gdbra).
If we shift $L\mapsto L + \lambda$ by a constant parameter, the Adler
map develops a linear term $J\mapsto J + \lambda J_\infty$, where
$$J_\infty(X) = -\comm{L}{X}_+\()$$
is now a differential operator of order at most $n-2$ and can
therefore be written as
$$J_\infty(X) = \sum_{i,j=0}^{n-2} \left( J^{(\infty)}_{ij}\cdot x_j
\right) \d^i~.\()$$
One can then define a Poisson bracket---the first Gel'fand--Dickey
bracket---by
$$\pb{u_i(x)}{u_j(y)}_\infty = -J^{(\infty)}_{ij}\cdot
\delta(x-y)~,\()$$
where, again, the differential operators $J^{(\infty)}_{ij}$ are taken
at the point $x$.  This is the first hamiltonian structure for the
isospectral flows of $L$.

Notice that $u_{n-1}$ is central relative to the first
Gel'fand--Dickey bracket, which implies that $u_{n-1}$ does not evolve
according to the Lax flows.  It is therefore consistent to impose the
constraint $u_{n-1}=0$.  This can be done for free in the first
Gel'fand--Dickey bracket, but it requires the Dirac bracket for the
second.  Hence we define
$$J^{(0)}_{ij} \equiv J_{ij} -
J_{i,n-1}\,J_{n-1,n-1}^{-1}\,J_{n-1,j}~,\()$$
which, despite the potential nonlocality present in
$J_{n-1,n-1}^{-1}$, is a differential operator.  It is the reduced
second Gel'fand--Dickey bracket,
$$\pb{u_i(x)}{u_j(y)}_0 = - J^{(0)}_{ij}\cdot \delta(x-y)~,\(Wn)$$
which provides a classical realization of $W_n$.  Indeed, for $n=2$,
$T=u_0$ obeys the Virasoro algebra; and, for $n=3$, $T=u_1$ and
$W=u_0-\fr 1/2 u_1'$ obey Zamolodchikov's $W_3$.  In general, the
$\{u_i\}$ can be redefined to $\{T,U_3,U_4,\ldots,U_n\}$ in such a way
that $T$ obeys the Virasoro algebra and each $U_i$ is a primary field
of weight $i$ \[DFIZ].
\section{Classical Limits}

The classical realization of $W_n$ defined by the reduced
second Gel'fand--Dickey bracket \(Wn) is encoded in the Adler map
\(Adler), which necessitates for its definition only the algebraic
structures present in the ring $R$ of pseudodifferential operators
(\pdo's).  The classical limit $w_n$ of $W_n$ will be obtained by
mimicking the definition of the Adler map in the classical limit of
$R$.  This ring $R$ is, in fact, an algebra and hence, in particular,
a vector space.  Taking the classical limit consists of endowing this
underlying vector space with the structure of a Poisson algebra, in
such a way that commutators go into Poisson brackets.  It is a general
fact (see, for example, \[KoSt]) that the classical limit exists
whenever, as in the case at hand, the ring $R$ is filtered---the
Poisson structure being defined on the associated graded object.
However, rather than appealing to the general theory, we will define
the classical limit directly and explicitly.

To every {\pdo} $A$ we associate its symbol---a formal Laurent
series---as follows.  We first write $A$ with all $\d$'s to the right:
$A = \sum_{i\leq N} a_i\d^i$.  (Each $A$ has a unique expression of
this form.)  Its symbol is then the formal Laurent series
in $\xi^{-1}$ given by
$$\widetilde{A} = \sum_{i\leq N} a_i\xi^i~.\()$$
Symbols have a commutative multiplication given by multiplying the
Laurent series; but one can define a composition law $\circ$ which
recovers the multiplication law in $R$.  In other words,
$$\widetilde{A}\circ\widetilde{B} = \widetilde{AB}~,\()$$
where $AB$ means the usual product of \pdo's.  Let $\widetilde{A}$ and
$\widetilde{B}$ be pseudodifferential symbols.  Then their composition is
easily shown to be given by
$$\widetilde{A}\circ\widetilde{B}= \sum_{k\geq 0} {1\over k!}
\kpder{k}{\widetilde{A}}{\xi} \kpder{k}{\widetilde{B}}{x}~.\()$$
For example, $\xi\circ a = a\xi + a'$ which recovers the basic Leibniz
rule: $\d a = a\d + a'$.

To define the classical limit we introduce a formal parameter
$\hbar$ in the composition law
$$\widetilde{A}\circ\widetilde{B}= \sum_{k\geq 0} {\hbar^k\over k!}
\kpder{k}{\widetilde{A}}{\xi} \kpder{k}{\widetilde{B}}{x}~.\(hbarcomp)$$
interpolating from the (commutative) multiplication of symbols for
$\hbar=0$ to the (noncommutative) multiplication of \pdo's for
$\hbar=1$.  The classical limit of any structure is obtained by
introducing the parameter $\hbar$ via \(hbarcomp) and keeping only the
lowest term in its $\hbar$ expansion.  Therefore, the classical limit
of $\circ$ is simply the commutative multiplication of symbols; hence
the name commutative limit.

The Poisson structure on symbols is defined as the classical limit of
the commutator---namely,\fnote{We use $\dlb{}{}$ to the denote the
Poisson bracket to avoid confusion with the Poisson bracket defining
the classical $W$-algebras.}
$$\dlb{\widetilde{A}}{\widetilde{B}} = \lim_{\hbar\to 0} \hbar^{-1}
\widetilde{\comm{A}{B}}~,\()$$
which can be written explicitly with the help of \(hbarcomp) as
$$\dlb{\widetilde{A}}{\widetilde{B}} = \pder{\widetilde{A}}{\xi}
\pder{\widetilde{B}}{x} -  \pder{\widetilde{A}}{x}
\pder{\widetilde{B}}{\xi}~.\()$$
One recognizes this at once as the standard Poisson bracket on a
two-di\-men\-sion\-al phase space with canonical coordinates $(x,\xi)$.

We must now take the classical limit of the Adler map \(Adler).  The
Adler map can be rewritten as follows:
$$J(X) = \comm{L}{X}_+ L - \comm{L}{(XL)_+}~,\()$$
which makes its classical limit obvious.  Indeed, if---dropping
tildes---$L = \xi^n + \sum_{i=0}^{n-1}u_i\xi^i$ and $X =
\sum_{i=0}^{n-1} x_i \xi^{-i-1}$ are symbols, the classical limit of
the Adler map $J(X)$ is given by
$$J_{c\ell}(X) = \dlb{L}{X}_+ L - \dlb{L}{(XL)_+} = \dlb{L}{(XL)_-} -
\dlb{L}{X}_-L~.\(Jcl)$$
It is clear from the above expression that $J_{c\ell}(X)$ is a
polynomial in $\xi$ of order at most $n-1$ and moreover that it is
linear in $X$.  This allows us to write
$$J_{c\ell}(X) = \sum_{i,j=0}^{n-1} \left( \Jt_{ij} \cdot
x_j\right) \xi^i~,\(Jclij)$$
where the $\Jt_{ij}$ are differential operators which are now at
most of order one.  The $\Jt_{ij}$ can be used to define
Poisson brackets
$$\pb{u_i(x)}{u_j(y)}^{c\ell} = - \Jt_{ij}\cdot
\delta(x-y)~,\(gdtoo)$$
as before.

It is easy to show that this bracket is indeed a Poisson bracket.  One
can see this from the fact that this is a contraction of the original
Adler map and that the Jacobi identity is satisfied order by order in
$\hbar$; or, alternatively, one can easily modify Dickey's proof
\[DickeyKW] of the Kupershmidt--Wilson theorem \[KW] to prove the
following.  If we formally factorize the symbol $L = \xi^n +
\sum_{i=0}^{n-1}u_i\xi^i = \prod_{i=1}^n (\xi + \phi_i)$, we find that
$u_i = \sigma_{n-i}(\phi)$---the elementary symmetric functions of the
$\{\phi_i\}$.  If we then define the Poisson bracket of the
$\{\phi_i\}$ as
$$\pb{\phi_i(x)}{\phi_j(y)} = \delta_{ij} \delta'(x-y)~,\(Miura)$$
and we compute the induced bracket on the $u_i$, we find precisely the
bracket \(gdtoo).

We can also find the classical limit of the first Gel'fand--Dickey
bracket by shifting $L \mapsto L + \lambda$ in the classical Adler map
$J_{c\ell}$ to obtain
$$J_{c\ell}^{(\infty)}(X) = -\dlb{L}{X}_+~.\(gdone)$$

To obtain the classical limit, $w_n$, of $W_n$ we perform the
reduction $u_{n-1}=0$ in the bracket \(gdtoo).  This defines
the $w_n$ algebra
$$\pb{u_i(x)}{u_j(y)}_0^{c\ell} = \Jt^{(0)}_{ij}\cdot
\delta(x-y)~,\(wn)$$
where $\Jt^{(0)}_{ij}$ are differential operators given by
$$\Jt^{(0)}_{ij} = \Jt_{ij} -
\Jt_{i,n-1}\Jt_{n-1,n-1}^{-1}\Jt_{n-1,j}~.\(Jcldirac)$$
\section{Explicit Structure of $w_n$}

We now compute the $w_n$ algebras explicitly.  Let $L = \xi^n +
\sum_{i=0}^{n-1} u_i\xi^i$ and $X=\sum_{i=0}^{n-1}x_i\xi^{-i-1}$ be
symbols.  The classical limit of the second Gel'fand--Dickey bracket
is given by equation \(gdtoo) where the differential operators
$\Jt_{ij}$ are given by \(Jcl) and \(Jclij).  A reasonably easy
computation yields the following $\Jt_{ij}$:
$$\veqnalign{\Jt_{n-1,n-1} &= -n\d~,\cr
\Jt_{i,n-1} &= -(i+1)u_{i+1}\d~,\cr
\Jt_{n-1,j} &= -(j+1)\d u_{j+1}~,\cr
\Jt_{ij} &= (n-j-1) \d u_{i+2+j-n} + (n-i-1) u_{i+2+j-n}\d\()\cr
&\quad{}+ \sum_{l=j+2}^{n-1} \left[ (l-i-1) u_{i+j+2-l}\d u_l +
(l-j-1) u_l \d u_{i+j+2-l}\right]\cr
&\quad{}-(i+1)u_{i+1}\d u_{j+1}~,\cr}$$
where $i,j=0,1,\ldots,n-2$ and with the proviso that $u_{l< 0}=0$ is
the above formulas.

After imposing the constraint $u_{n-1}=0$, the induced bracket is
given by \(wn) where the differential operators $\Jt^{(0)}_{ij}$ are
given by \(Jcldirac).  An easy computation yields the following
$\Jt^{(0)}_{ij}$:
$$\veqnalign{\Jt^{(0)}_{ij} &= (n-j-1)\d u_{i+2+j-n} + (n-i-1)
u_{i+2+j-n}\d \cr
&\quad{}+ \sum_{l=j+2}^{n-1} \left[ (l-i-1) u_{i+j+2-l}\d u_l +
(l-j-1) u_l \d u_{i+j+2-l}\right]\cr
&\quad{}+\fr{(i+1)(j+1-n)}/n u_{i+1}\d u_{j+1}~,\(explwn)\cr}$$
where now both $i,j=0,1,\ldots,n-2$ and with the proviso that
$u_{n-1}=u_{l<0}=0$.  In particular, if $j=n-2$ we find
$$\Jt^{(0)}_{i,n-2} = (n-i)u_i\d + u_i'~,\()$$
which identifies $u_{n-2}$ as a generator of a $\Diff\,S^1$ subalgebra
and each $u_i$ as a $\Diff\,S^1$ tensor of weight $n-i$.  Therefore,
the natural basis for $w_n$---namely, the coefficients in the Lax
symbol---are already primary fields.

We can easily work out the first two examples $n=2$ and $n=3$.  For
$n=2$ we only have one field, $T=u_0$, and the algebra is clearly
$\Diff\,S^1$:
$$\pb{T(x)}{T(y)}_0^{c\ell} = - \left[ 2 T(x)\d + T'(x)\right] \cdot
\delta(x-y)~.\(diff)$$
For $n=3$ we define $T=u_1$ and $W=u_0$.  Then from \(explwn) we find
\(diff) together with
$$\veqnalign{\pb{W(x)}{T(y)}_0^{c\ell} &= - \left[ 3 W(x)\d +
W'(x)\right] \cdot \delta(x-y)~,\cr
\noalign{\hbox{and}}
\pb{W(x)}{W(y)}_0^{c\ell} &= - \left[ {2\over3} T(x)\d T(x)\right]
\cdot \delta(x-y)~,\(wthree)\cr}$$
which is $w_3$.
\section{Reduction from $w_{\rm KP}$ to $w_n$}

The space of Lax operators $L =\d^n + \sum_{i=0}^{n-1} u_i\d^i$ is
naturally a subspace of the space $M^{(n)}$ of all \pdo's of the form
$\d^n + \sum_{i\leq n-1}u_i\d^i$.  The Adler map \(Adler) naturally
extends to $M^{(n)}$ by the same formula, but where $L$ is now
understood as a {\pdo} in $M^{(n)}$ and $X = \sum_{i\leq n-1}
\d^{-i-1}x_i$.  The expression for $J(X)$ tells us that it is a
{\pdo} of order at most $n-1$, whence we can write it as
$$J(X) = \sum_{i,j\leq n-1} \left( \Omega_{ij}\cdot x_j \right)
\d^i~,\()$$
where the $\Omega_{ij}$ are differential operators defining a Poisson
bracket as in \(gdbra).  For reasons that will appear clear below, we
call this resulting algebra $W_{\rm KP}^{(n)}$.

The subspace of $M^{(n)}$ obtained by demanding that $L$ be
differential inherits a Poisson bracket which coincides with the one
derived from the original Adler map \(Adler).  In terms of hamiltonian
reduction, the constraints can be written simply as $L_-=0$.  Moreover
the constraint $u_{n-1}=0$ can be imposed before or after the
reduction from $M^{(n)}$.  In other words, $W_n$ is a hamiltonian
reduction of $W_{\rm KP}^{(n)}$. Taking the classical limit of the
previous sentence, we find that $w_n$ is a hamiltonian reduction of
the classical limit of $W_{\rm KP}^{(n)}$.  We will see below that
this limit is independent of $n$.

Every operator $L$ in $M^{(n)}$ has a unique $n^\th$ root of the form
$\Lambda = \d + \sum_{i\geq 0} a_i\d^{-i}$---that is, a KP
operator---and, similarly, every KP operator $\Lambda$ can be raised
to the $n^\th$ power to yield a {\pdo} in $M^{(n)}$.  Therefore the
space $M^{(n)}$ is isomorphic to the space $M^{(1)}$ of KP operators.
It was show in \[WKP] that in $M^{(1)}$ one can define a Poisson
bracket from a generalization of the Adler map
$$J_{\rm KP}(X) = (\Lambda X)_+\Lambda - \Lambda(X\Lambda)_+ =
\Lambda(X\Lambda)_- -(\Lambda X)_- \Lambda~,\(KPAdler)$$
for $\Lambda\in M^{(1)}$ and $X = \sum_{i\geq 0}\d^{i-1}x_i$.  The
associated Poisson bracket defines a $W$--algebra, called $W_{\rm
KP}$, which was shown to be a hamiltonian structure for the KP
hierarchy.

Now, it can be shown that $W_{\rm KP}^{(n)}$---the extension of the
Adler map to $M^{(n)}$---is also a hamiltonian structure for the
KP hierarchy, which explains the name.  It can be shown by explicit
computation of the first few Poisson brackets that these algebras are
all different and, in particular, different from $W_{\rm KP}$.
Nevertheless, as we now show, they all have the same classical limit.

The algebra $w_{\rm KP}$ is the Poisson algebra induced by the
classical limit of the generalized Adler map \(KPAdler) which can be
written as
$$J_{\rm KP}^{c\ell}(X) = \dlb{\Lambda}{X}_+ \Lambda -
\dlb{\Lambda}{(X\Lambda)_+}~,\(wKP)$$
where $\Lambda = \xi + \sum_{i\geq 0} a_i \xi^{-i}$ and $X =
\sum_{i\geq 0} x_i\xi^{i-1}$ are symbols.  In the usual manner this
defines a Poisson bracket on the variables $\{a_i\}$ which defines the
$w_{\rm KP}$ algebra.  Now let $\varphi$ denote the map sending
$\Lambda$ to its $n^\th$ power $\Lambda^n = \xi^n + \sum_{i\leq n-1}
u_i\xi^i$.  The $\{u_i\}$ can then be solved as polynomials in the $\{a_i\}$
and this map is in fact invertible, so that the $\{a_i\}$ can be in turn
be solved for as polynomials in the $\{u_i\}$.  The Poisson bracket of the
$\{a_i\}$ then induces a Poisson bracket among the $\{u_i\}$ which defines a
classical $W$--algebra.  On the other hand, the $\{u_i\}$ inherit another
Poisson bracket from the classical limit of the extension of the Adler
map to $M^{(n)}$ and it is this bracket which we have seen to reduce
to $w_n$.  What we will now show is that these two brackets on the
$\{u_i\}$ are in fact the same.

This uses some geometric formalism which is reviewed, for example, in
\[Laxform].  Roughly the Adler map can be understood as a tensorial
map from 1-forms to vector fields.  The map $\varphi: M^{(1)} \to
M^{(n)}$ then induces an Adler-type map on $M^{(n)}$ from the one on
$M^{(1)}$ as follows.  If $X$ is any 1-form on $M^{(n)}$, we first
pull it back to $M^{(1)}$ via $\varphi^*$.  We then apply the map
\(KPAdler) to it yielding a vector field on $M^{(1)}$, which can be
then pushed forward to $M^{(n)}$ with $\varphi_*$. Therefore the
induced Poisson brackets will be the ones associated to the Adler-type
map $\varphi_* \circ J_{\rm KP}^{c\ell} \circ \varphi^*$.  We now
proceed to compute this.  A straightforward calculation shows that if
$A$ is a tangent vector on $M^{(1)}$ at $\Lambda$, its push-forward is
the tangent vector to $M^{(n)}$ at $L=\Lambda^n$ given by $\varphi_* A
= n \Lambda^{n-1}A$.  Similarly, if $X$ is a 1-form on $M^{(n)}$ at
$L$, its pull-back is the 1-form on $M^{(1)}$ at $L^{1/n} = \Lambda$
given by $\varphi^* X = n \Lambda^{n-1}X$.  Therefore, the induced
hamiltonian map is given by
$$\veqnalign{\left( \varphi_* \circ J_{\rm KP}^{c\ell} \circ \varphi^*
\right)(X) &= n^2 \Lambda^{n-1} \left(
\dlb{\Lambda}{\Lambda^{n-1}X}_+\Lambda - \dlb{\Lambda}{(\Lambda^n
X)_+}\right)\cr
&= n L \dlb{L}{X}_+ - n \dlb{L}{(LX)_+}~,\(Induced)\cr}$$
where we have used repeatedly the fact that
$n\dlb{\Lambda}{\Lambda^{n-1}Z} = n \dlb{\Lambda}{Z}\Lambda^{n-1} =
\dlb{L}{Z}$ for any $Z$.  But---up to the inessential multiplicative
factor $n$---we recognize in the right-hand side of \(Induced) the
classical limit of the extension to $M^{(n)}$ of the Adler map
\(Adler).  This concludes the proof.
\section{A Word on Universality}

Reductions clearly commute with classical limits.  From this we
conclude that since $W_n$ is a reduction of $W_{\rm KP}^{(n)}$,
$w_n$---the classical limit of $W_n$---is a reduction of the classical
limit of $W_{\rm KP}^{(n)}$, \ie, $w_{\rm KP}$.  But this limit is
$n$-independent, hence all $w_n$ are hamiltonian reductions of $w_{\rm
KP}$.

In \[Uniw] we proved, as an easy corollary of the Kupershmidt-Wilson
theorem, that $W_n$ is a hamiltonian reduction of $W_{n+1}$ for any
$n$.  This allowed us to define a universal $W_n$-algebra as the
inverse limit of the $W_n$ under these reductions.  Existence and
uniqueness then followed from the universal properties of (co)limits.
One can repeat the steps in \[Uniw] for the $w_n$ algebras, to show
that $w_n$ is a hamiltonian reduction of $w_{n+1}$ for every $n$ and
therefore define a universal algebra as their colimit.  Roughly, this
algebra has the property that all $w_n$ algebras can be obtained as
hamiltonian reductions from it and that, in some sense, it is the
smallest such algebra.  The remarks at the beginning of this section
thus clearly suggest that $w_{\rm KP}$ is the universal (classical)
$W$--algebra for the $w_n$ series.

Moreover, we expect it to be the classical limit of the universal
$W$--algebra for the $W_n$ series.  Or, said differently, that the
universal $W$--algebra of \[Uniw] is a deformation of $w_{\rm KP}$.
We have already seen that $w_{\rm KP}$ has an infinite number of
different deformations $W_{\rm KP}^{(n)}$.  Since the universal
$W$-algebra is unique, it would have to be at most one of them, and
the ``principle of insufficient reason'' would dictate that if it is
one of them, it should be $W_{\rm KP}$.  However, we have not yet been
able to exhibit any $W_n$ as a hamiltonian reduction of $W_{\rm KP}$.
\section{Deformation to $w_{1+\infty}$}

Shifting $\Lambda \mapsto \Lambda + \lambda$ in \(wKP) defines another
hamiltonian map
$$J^{c\ell}_{{\rm KP},\infty}(X) = -
\dlb{\Lambda}{X_+}_-~,\(FirstStruc)$$
which defines fundamental Poisson brackets for the coefficients
$\{a_i\}$ of $\Lambda$.  In fact, if we denote by
$\Omega^{(\infty)}_{ij}$ the differential operators defined by
$$J^{c\ell}_{{\rm KP},\infty}(X) = \sum_{i,j=0}^\infty \left(
\Omega^{(\infty)}_{ij}\cdot x_j \right) \xi^{-i}~,\()$$
for $X = \sum_{i=0}^\infty x_i\xi^i$, the fundamental Poisson bracket
of the $\{a_i\}$ associated to the hamiltonian map \(FirstStruc) is
given by
$$\pb{a_i(x)}{a_j(y)}^{c\ell}_{{\rm KP},\infty} = -
\Omega^{(\infty)}_{ij}\cdot \delta(x-y)~.\()$$

These differential operators can be computed explicitly.  Notice that
since only $X_+$ appears, $x_0$ never enters the picture, whence
$\Omega^{(\infty)}_{0j}= \Omega^{(\infty)}_{i0}=0$.  Therefore $a_0$
is central and can be dropped without harm.  For $i,j\geq 1$ one finds
$$ \Omega^{(\infty)}_{ij} = (j-1)\d a_{i+j-2} + (i-1) a_{i+j-2}\d
{}~.\(winfty)$$
Notice that $\Omega^{(\infty)}_{11}=0$ and that
$$\Omega^{(\infty)}_{22} = \d a_2 + a_2\d~,\()$$
which means that $a_2$ generates a $\Diff\,S^1$ subalgebra.  Moreover,
$$\Omega^{(\infty)}_{i2} =  a_i' + i a_i\d~,\()$$
which identifies $a_i$ as a $\Diff\,S^1$ tensor of weight $i$.  In
fact, \(winfty) is nothing but the algebra $w_{1+\infty}$ of
hamiltonian vector fields on the 2-plane, with coordinates $(x,y)$,
which depend polynomially on $y$ \[Bakas]\[Pope].  Also, the
subalgebra generated by $\{a_i\}_{i\geq 2}$ is nothing but $w_\infty$.
Therefore we see that $w_{\rm KP}$ deforms to $w_{1+\infty}$.
\section{Conclusions and Outlook}

In this paper we have studied the classical limits of the $W_n$
algebras.  We call these algebras $w_n$ and they are nonlinear
extensions of the diffeomorphism algebra of the circle, $\Diff\,S^1$,
by tensors of weights $3,4,\ldots,n$.  Their construction followed the
construction of the realization of $W_n$ as the second hamiltonian
structure of the generalized KdV hierarchy.  These realizations are
defined in terms of the algebraic structure of the ring of
pseudodifferential operators and therefore their classical limit is
simply obtained by repeating the construction with the commutative
limit of this ring: the Poisson algebra of pseudodifferential symbols,
which is isomorphic to the Poisson algebra of smooth sections of the
cotangent bundle (with the zero section removed) of the circle under
the canonical symplectic structure.

These algebras are simplified versions of the $W$-algebras appearing
in the Lax formalism; but they contain their essential feature, namely
polynomial nonlinearity.  We think that they make suitable candidates
to begin to understand things $W$, in particular $W$-geometry.

The results of the last section on the reduction from $w_{\rm KP}$ to
$w_n$ beg for the development of the deformation theory of
$W$-algebras.  In the context in which they are treated here, they are
particular examples of Poisson algebras, whereas in the quantum case
they are particular examples of associative algebras, both of which
count with a reasonably well-developed deformation theory.  But it
would be interesting to see if the features which characterize
$W$-algebras (both classically among Poisson algebras and
quantum-mechanically among associative algebras) allow for some
simplification of their deformation theory which would make at least
their infinitesimal deformations readily computable.

In this letter we have been rather sketchy in the development of the
formalism.  The details as well as many other results will appear in
our forthcoming paper \[Next], which will also treat the
supersymmetric case.  There are other issues that deserve some study,
\eg, uniqueness of classical $W$-algebras, relation with Lie algebras
and with integrable systems, quantization, \dots which will be treated
in \[Next] and/or in future work.

\ack

We would like to thank Javier Mas for pointing reference \[TakaTake]
to us and for his interest.  We are also grateful to Eva R{\'\i}os for
insightful discussions.
\refsout
\bye